\documentclass[
aps,%
12pt,%
final,%
notitlepage,%
oneside,%
onecolumn,%
nobibnotes,%
nofootinbib,%
superscriptaddress,%
noshowpacs,%
centertags]%
{revtex4}

\usepackage{bm}
\usepackage{graphics}
\usepackage{rotating}
\usepackage{amsmath}
\usepackage{amssymb}
\usepackage{mathtext}
\usepackage{upgreek}

\usepackage{epsfig}

\usepackage{tabularx}

\begin{document}

\title{Paired charmonium and bottomonium production in rare\\ exclusive decays of $Z$ boson}

\author{\firstname{F.~A.} \surname{Martynenko}}
\email{f.a.martynenko@gmail.com}
\affiliation{Samara University, Samara, Russia}

\author{\firstname{A.~P.} \surname{Martynenko}}
\affiliation{Samara University, Samara, Russia}

\author{\firstname{A.~A.} \surname{Skakun}}
\affiliation{Samara University, Samara, Russia}

\begin{abstract}
The processes of paired charmonium and bottomonium production in $Z$-boson decays are investigated 
within the relativistic quark model. Various decay mechanisms are examined, and relativistic decay amplitudes 
are constructed, taking into account the relative momenta of heavy quarks. Decay widths are calculated 
for various mechanisms in the nonrelativistic approximation and taking into account relativistic corrections.
\end{abstract}

\maketitle

\section{Introduction}

Following the discovery of the Higgs boson, research into various processes in the Higgs sector 
of the Standard Model has increased significantly. This research is largely aimed at more accurately 
determining the model's parameters, including the numerous interaction constants of various particles 
and their masses. An important part of the reactions being studied involves the production of quark 
and lepton bound states. These reactions allow for testing various models of quarkonia and leptonium 
production and searching for as-yet-undiscovered states.
Among such reactions, reactions with exclusive and inclusive production of a pair of bound states 
of particles can be distinguished, 
since in such reactions the effects of particle binding are manifested much more strongly and testing 
the theory of the formation of ordinary mesons or baryons can be more successful 
\cite{bander,keung,kln,p3,apm1,gonsalves,kniehl,sun}.

In recent years, experimental studies of the pair production of heavy quarkonia in Higgs and $Z$ boson 
decays have intensified. So far, these have only resulted in upper bounds on the possible decay widths
\cite{CMS2023,de1,enterria}:
\begin{equation}
\label{eq1}
{\cal B}_{Z\to J/\Psi+J/\Psi}\leq 1.1\cdot 10^{-6},~~~{\cal B}_{Z\to \Upsilon(mS)+\Upsilon(nS)}\leq 3.9\cdot 10^{-7},~~~
{\cal B}_{Z\to \Upsilon(1S)+\Upsilon(1S)}\leq 1.8\cdot 10^{-6},
\end{equation}
\begin{equation}
\label{eq2}
{\cal B}_{H\to J/\Psi+J/\Psi}< 3.8\cdot 10^{-4},~~~{\cal B}_{H\to \Upsilon(1S)+\Upsilon(1S)}\leq 1.7\cdot 10^{-3}.
\end{equation}

Although charmonium pair production has been studied by various authors both within a nonrelativistic approach 
and taking into account radiative and relativistic effects \cite{p1,p2,p3,p4,p5,p6}, 
approaches to calculating these processes continue 
to improve. Nonrelativistic quantum chromodynamics methods and various quark models based on the Bethe-Salpeter 
method or the quasipotential approach are used to calculate the observed decay 
widths and production rates \cite{nrqcd,akl1,akl2,ebert,bl,QWG,Fan}. 
Along with charmonium 
pair production, the pair production of upsilon mesons or mixed pair production of mesons consisting of $c$ 
and $b$ quarks is also considered \cite{p7,p8}.

Improving the calculation of decay widths for such processes involves considering different decay mechanisms. 
The key parameters ultimately determining numerical value of decay width are the constants 
of electromagnetic and strong interactions, the Weinberg angle, and the ratio of masses 
of the produced mesons to the mass of the $Z$-boson. In a certain sense, the mass parameter is also 
key, since the significance of a specific decay mechanism depends on it.

Our previous studies of processes in the Higgs sector focused on Higgs boson decay \cite{apm1,apm2,apm3}. 
In this paper, we extend 
the scope of our study of pair quarkonium production processes to include $Z$-boson decays. The primary goal of this 
study is to calculate relativistic effects within the method we are developing, 
based on the relativistic quark model. Furthermore, we explore various decay mechanisms to determine 
the mechanism that provides the leading-order contribution in terms of parameters
governing the decay processes.

\section{General formalism}

There are various mechanisms for the decay of the $Z$-boson, describing the pair production of charmonium or
bottomonium. The Feynman diagrams in Fig.~\ref{Z-QG} show the decay amplitudes that we attribute 
to the quark-gluon mechanism. In this case, 
the first perturbative stage of the process involves the production of a heavy quark and an antiquark, the emission 
of a gluon, and the production of a second quark-antiquark pair. The second stage of the process involves 
the nonperturbative formation of charmonium (bottomonium) from free quark-antiquark pairs.
The vertex function describing the transformation of the $Z$-boson into a quark-antiquark pair 
has the form \cite{brs,pp1}:

\begin{figure}[htbp]
\centering
\includegraphics[scale=1.]{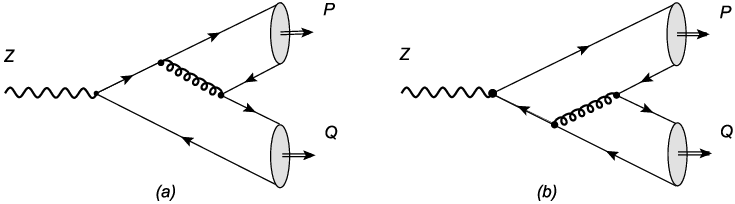}
\caption{Quark--gluon mechanism of charmonium (bottomonium) pair production. $P$, $Q$ are four-momenta 
of heavy quarkonium.}
\label{Z-QG}
\end{figure}

\begin{equation}
\label{eq3}
\hat \Gamma=\varepsilon_\alpha\Gamma^\alpha,~~~
\Gamma^\alpha=\frac{e}{\sin 2\theta_{\rm W}} \gamma^\alpha
\left[\frac{1}{2}(1-\gamma_5)-a_q\right],~~~a_q=2Q_q\sin^2\theta_{\rm W},
\end{equation}
where $M_Z$ is the $Z$-boson mass, $Q_q$ is the $q$-quark charge ($Q_c$=2/3, $Q_b$=-1/3), 
$\theta_{\rm W}$ is the 
Weinberg angle, $\varepsilon^\alpha$ is the polarization vector of $Z$-boson.

We consider the production of $S$-wave vector and pseudoscalar states of charmonium (bottomonium).
In the rest frame of the $Z$-boson, the produced states of heavy quarkonium move with four-momenta $P$ and $Q$.
In the relativistic quasipotential quark model there are four decay amplitudes which
can be presented as a convolution of a perturbative production amplitude
of two $c(b)$-quarks and two $\bar c(\bar b)$-antiquarks and quasipotential
wave functions of final mesons \cite{apm1,apm2}:
\begin{footnotesize}
\begin{equation}
\label{eq4}
{\cal M}^{(1)}(P,Q)=\frac{16\pi e\alpha_s}{3\sin 2\theta_W}\int\frac{d{\bf p}}{(2\pi)^3}\int\frac{d{\bf q}}{(2\pi)^3}
Tr\left\{\Psi^{\cal V}(p,P)\gamma^\nu S_c(r-q_2)\hat\Gamma\Psi^{\cal P}(q,Q)\gamma^\lambda\right\}
D_{\nu\lambda}(p_2+q_1),
\end{equation}
\begin{equation}
\label{eq5}
{\cal M}^{(2)}(P,Q)=\frac{16\pi e\alpha_s}{3\sin 2\theta_W}\int\frac{d{\bf p}}{(2\pi)^3}\int\frac{d{\bf q}}{(2\pi)^3}
Tr\left\{\Psi^{\cal V}(p,P)\gamma^\nu S_c(p_1-r)\hat\Gamma\Psi^{\cal P}(q,Q)\gamma^\lambda\right\}
D_{\nu\lambda}(p_2+q_1),
\end{equation}
\begin{equation}
\label{eq6}
{\cal M}^{(3)}(P,Q)=\frac{16\pi e\alpha_s}{3\sin 2\theta_W}\int\frac{d{\bf p}}{(2\pi)^3}\int\frac{d{\bf q}}{(2\pi)^3}
Tr\left\{\Psi^{\cal P}(q,Q)\gamma^\nu S_c(r-p_2)\hat\Gamma\Psi^{\cal V}(p,P)\gamma^\lambda\right\}
D_{\nu\lambda}(p_1+q_2),
\end{equation}
\begin{equation}
\label{eq7}
{\cal M}^{(4)}(P,Q)=\frac{16\pi e\alpha_s}{3\sin 2\theta_W}\int\frac{d{\bf p}}{(2\pi)^3}\int\frac{d{\bf q}}{(2\pi)^3}
Tr\left\{\Psi^{\cal P}(q,Q)\gamma^\nu S_c(q_1-r)\hat\Gamma\Psi^{\cal V}(p,P)\gamma^\lambda\right\}
D_{\nu\lambda}(p_1+q_2),
\end{equation}
\end{footnotesize}
where the symbol hat denotes convolution of four-vector with the Dirac $\gamma$-matrices.
The overall color factor $T^a_{ij}\frac{1}{\sqrt{3}}\delta_{jk}T^a_{kl}\frac{1}{\sqrt{3}}\delta_{li}=\frac{4}{3}$ 
is taken into account in \eqref{eq4}-\eqref{eq7}.
Four-momenta $p_1$ and $p_2$ of $c(b)$-quark and $\bar c(\bar b)$-antiquark in the pair
forming the first $(c\bar c)$ or $(b\bar b)$ meson, and four-momenta $q_1$ and $q_2$ for
quark and antiquark in the second meson are expressed in terms of relative and total 
four-momenta as follows:
\begin{equation}
\label{eq8}
p_{1,2}=\frac12 P \pm p,\quad (pP)=0; \qquad q_{1,2}=\frac12 Q \pm q,\quad (qQ)=0,
\end{equation}
A superscript ${\cal V}$ indicates a vector meson $ (c\bar c)$ ($ (b\bar b)$) and
a superscript ${\cal P}$ indicates a pseudoscalar meson $(c\bar c)$ ($ (b\bar b)$).
$S_c(p)$ is the heavy quark propagator, $D^{\lambda\sigma}(k)$ is the gluon propagator.

Heavy quarks $c$, $b$ and antiquarks $\bar c$, $ \bar b$ are outside 
the mass shell in the intermediate state:
$p_{1,2}^2\not= m^2$, so that $p_1^2-m^2=p_2^2-m^2$, 
which means that there is symmetrical exit of particles from the mass shell.

Relativistic wave functions of pseudoscalar $\Psi_{\cal P}$ and vector 
$\Psi_{\cal V}$ bound states of quarks have the form \cite{apm1,apm2,apm3}:
\begin{eqnarray}
\label{eq9}
\Psi^{\cal P}(p,P)&=&\frac{\Psi({\bf p})}{[\frac{\epsilon(p)}{m}\frac{(\epsilon(p)+m)}{2m}]}
\left[\frac{\hat v_1-1}{2}-\hat
v_1\frac{{p}^2}{2m(\epsilon(p)+ m)}-\frac{\hat{p}}{2m}\right] \times \cr
&&\frac{\gamma_5(1+\hat v_1)}{2\sqrt{2}}  \left[\frac{\hat
v_1+1}{2}-\hat v_1\frac{{p}^2}{2m(\epsilon(p)+m)}+\frac{\hat{p}}{2m}\right],
\end{eqnarray}
\begin{eqnarray}
\label{eq10}
\Psi^{\cal V}(q,Q)&=&\frac{\Psi({\bf q})}
{[\frac{\epsilon(q)}{m}\frac{(\epsilon(q)+m)}{2m}}
\left[\frac{\hat v_2-1}{2}-\hat v_2\frac{{q}^2}{2m(\epsilon(q)+
m)}+\frac{\hat{q}}{2m}\right] \times \cr 
&&\frac{\hat{\varepsilon}(Q,S_z)(1+\hat v_2)}{2\sqrt{2}} 
\left[\frac{\hat v_2+1}{2}-\hat v_2\frac{{q}^2}{2m(\epsilon(q)+ m)}-\frac{\hat{q}}{2m}\right],
\end{eqnarray}
$v_1=P/M_1$, $v_2=Q/M_2$, $M_{1,2}$ are the meson masses, $m$ is the $c(b)$-quark mass,
$\varepsilon^\lambda(Q,S_z)$ is the charmonium (bottomonium) spin four-vector,
$\epsilon(p)=\sqrt{p^2+m^2}$ is relativistic energy of quarks.
$\Psi({\bf p})$ is the charmonium (bottomonium) wave function in the rest frame.
Relative momenta $p=L_P(0,{\bf p})$ and
$q=L_Q(0,{\bf q})$ are obtained after the Lorentz transformation of four-vectors $(0,{\bf p})$ and
$(0,{\bf q})$ to the reference frames moving with four-momenta $P$ and $Q$.

We have omitted here intermediate expressions, leading to equations~\eqref{eq4}-\eqref{eq7}
because they were discussed in detail in our previous papers \cite{EFGM2009,apm2005}.
In the Bethe-Salpeter approach the initial production amplitude has a form of convolution of the truncated
amplitude with two Bethe-Salpeter (BS) charmonium wave functions.
The presence of the $\delta (p\cdot P)$-function in this case
allows us to make the integration over relative energy $p^0$. In the rest frame of a bound state the condition
$p^0=0$ allows to eliminate the relative energy from the BS wave function.

Relativistic wave functions in Eqs.~\eqref{eq9} and \eqref{eq10} are the product
of meson wave functions in the rest frame $\Psi_0({\bf p})$ and spin projection operators that are
accurate at all orders in $|{\bf p}|/m$ \cite{apm1,apm2}. 
Spin projection operators in \eqref{eq9}-\eqref{eq10} can be considered as form factors 
for the transition of quarks from a free to a bound state.
An expression of spin projector in different
form has been derived primarily in \cite{bodwin2002} where spin projectors are
written in terms of heavy quark momenta $p_{1,2}$ lying on the mass shell.
The transformation law of bound state wave function from the
rest frame to the moving one with four-momentum $P$
was discussed in the Bethe-Salpeter approach in \cite{brodsky} and in quasipotential method in \cite{faustov}. 
We use the quasipotential approach and write the transformation law of bound state wave function 
as follows \cite{faustov,apm1,apm2,apmsymmetry}:
\begin{equation}
\label{eq11}
\Psi_{P}^{\rho\omega}({\bf p})=D_1^{1/2,~\rho\alpha}(R^W_{L_{P}})
D_2^{1/2,~\omega\beta}(R^W_{L_{P}})\Psi_{0}^{\alpha\beta}({\bf p}),
\end{equation}
\begin{displaymath}
\bar\Psi_{P}^{\lambda\sigma}({\bf p})
=\bar\Psi^{\varepsilon\tau}_{0}({\bf p})D_1^{+~1/2,~\varepsilon
\lambda}(R^W_{L_{P}})D_2^{+~1/2,~\tau\sigma}(R^W_{L_{P}}),
\end{displaymath}
where $R^W$ is the Wigner rotation, $L_{P}$ is the Lorentz boost from the meson rest frame to a moving 
one, and the rotation matrix $D^{1/2}(R)$ is defined by the formula \cite{apm1,faustov}:
\begin{equation}
\label{eq12}
{1 \ \ \,0\choose 0 \ \ \,1}D^{1/2}_{1,2}(R^W_{L_{P}})=
S^{-1}({\bf p}_{1,2})S({\bf P})S({\bf p}),
\end{equation}
where explicit form for the Lorentz transformation matrix of the four-spinor is the following:
\begin{equation}
\label{eq13}
S({\bf p})=\sqrt{\frac{\epsilon(p)+m}{2m}}\left(1+\frac{(\boldsymbol{\alpha}{\bf p})} 
{\epsilon(p)+m}\right).
\end{equation}
For further transformation of initial expression of the amplitude the following relations
are applied \cite{faustov,apm1,apmsymmetry}:
\begin{equation}
\label{eq14}
S_{\alpha\beta}(\Lambda)u^\lambda_\beta(p)=\sum_{\sigma=\pm 1/2}
u^{\sigma}_\alpha(\Lambda p)D^{1/2}_{\sigma\lambda}(R^W_{\Lambda p}),
\end{equation}
\begin{displaymath}
\bar u^\lambda_\beta(p)S^{-1}_{\beta\alpha}(\Lambda)=\sum_{\sigma=\pm 1/2}
D^{+~1/2}_{\lambda\sigma}(R^W_{\Lambda p})\bar u^\sigma_\alpha(\Lambda p).
\end{displaymath}

Relativistic wave functions \eqref{eq9}-\eqref{eq10} contain spin projection operators onto states 
with spin 1 and 0 in the rest system:
\begin{equation}
\label{eq15}
\hat\Pi_{S=1}=[v(0)\bar u(0)]_{S=1}=\frac{\hat\varepsilon_1(1+\hat v_1)}{2\sqrt{2}},~~~
\hat\Pi_{S=0}=[v(0)\bar u(0)]_{S=0}=\frac{\gamma_5(1+\hat v_1)}{2\sqrt{2}},
\end{equation}
where $\varepsilon_1^\alpha$ is the polarization vector of $J/\Psi$ ($\Upsilon$) meson.

The first simplification of these amplitudes is related to denominators of quark and gluon 
propagators, where the mass of the $Z$-boson arises. Therefore, neglecting corrections 
of the $p^2/M_Z^2$, $q^2/M_Z^2$ type, we obtain:
\begin{equation}
\label{eq16}
\frac{1}{(r-p_{1,2})^2-m^2}\approx\frac{1}{r^2-rP}\approx \frac{2}{M_Z^2},~~~
\frac{1}{(p_{2,1}+q_{1,2})^2}\approx \frac{4}{M_Z^2}.
\end{equation}

Calculating the trace of a product of the Dirac factors in the Form package \cite{form}, 
we extract in the numerator second-order relativistic corrections $p^2/m^2$, $q^2/m^2$ 
in relative momenta $p$ and $q$, using averaging over the angles:
\begin{equation}
\label{eq17}
\langle p^\alpha p^\beta\rangle = \frac{{\bf p}^2}{3}\left(-g^{\alpha\beta}+v_1^\alpha v_1^\beta\right),~~~
\langle q^\alpha q^\beta\rangle = \frac{{\bf q}^2}{3}\left(-g^{\alpha\beta}+v_2^\alpha v_2^\beta\right).
\end{equation}

In the case of quark-gluon mechanism of pair quarkonium production, the numerator of total amplitude 
\eqref{eq4}-\eqref{eq7} has the form:
\begin{equation}
\label{eq18}
{\cal N}_{QG}=\varepsilon_{\alpha\beta\lambda\sigma}v_1^\alpha v_2^\beta
\varepsilon_1^\lambda\varepsilon^\sigma M F_{QG},
\end{equation}
\begin{displaymath}
F_{QG}=(1-2a_c)(\frac{1}{2}-2r_2-\frac{1}{2}r_1-\frac{1}{2}\omega_{1q}-2r_2\omega_{1q}+
\frac{1}{2}r_1\omega_{1q}-\frac{1}{6}\omega_{1p}+\frac{2}{3}r_2\omega_{1p}-\frac{1}{6}r_1\omega_{1p}),
\end{displaymath}
where the first part of the introduced parameters is determined by the ratio of particle masses:
($M_1=M_{J/\Psi}$ or $M_{\Upsilon}$, $M_2=M_{\eta_c}$ or $M_{\eta_b}$, $M=2m$):
\begin{equation}
\label{eq19}
r_1=\frac{M_1}{M},~~~r_2=\frac{M_2}{M},~~~r_3=\frac{M_Z}{M}.
\end{equation}

The other part of parameters $\omega_{1q}$, $\omega_{1p}$ describe the second-order relativistic corrections, 
which in the case of $S$-states are determined by the following convergent momentum integrals $I_{n}$:
\begin{equation}
\label{eq20}
I_{n}=\sqrt{\frac{2}{\pi}} \int_0^\infty p^2R(p)\frac{(\epsilon(p)+m)}{2\epsilon(p)}
\left(\frac{\epsilon(p)-m}{\epsilon(p)+m}\right)^n dp,~~~
\omega_{1}=\frac{I_{1}}{I_{0}},~\omega_{2}=\frac{I_{2}}{I_{0}},
\end{equation}
where $R(p)$ is radial wave function of the charmonium (bottomonium),
\begin{equation}
\label{eq21}
I_{0} = \tilde R(0)=\sqrt{\frac{2}{\pi}}\int_0^\infty \frac{(\epsilon(p)+m)}{2\epsilon(p)}p^2R(p)dp.
\end{equation}

Since the values of these parameters differ slightly for vector and pseudoscalar mesons, additional 
indices $p$ and $q$ are introduced for them. The index $p$ hereafter refers to a vector meson, 
and the index $q$ to a pseudoscalar meson.
The approach based on the relativistic quark model makes it possible to calculate all the introduced parameters.

The differential width of the $Z$-boson decay into a pair of charmonia $J/\Psi+\eta_c$ 
(similarly for a pair $\Upsilon+\eta_b$ ) is determined by the formula:
\begin{equation}
\label{eq22}
d\Gamma=\frac{|{\bf P}|}{32\pi^2 M_Z^2} \overline{|{\cal M}(Z\to J/\Psi + \eta_c)|^2}d\Omega,
\end{equation}
where the modulus of a charmonium (bottomonium) momentum vector has the form:
\begin{equation}
\label{eq23}
|{\bf P}|=\frac{1}{2M_Z}\sqrt{[M_Z^2-(M_1-M_2)^2][M_Z^2-(M_1+M_2)^2]}.
\end{equation}

Summing over polarizations of vector quarkonium
and averaging over polarizations of the $Z$-boson,
we obtain total decay width $Z\to V+P$ for the quark-gluon mechanism 
with the account of relativistic corrections as follows:
\begin{equation}
\label{eq24}
\Gamma(Z\to V+P) = 
\frac{2^{13}\pi^2\alpha\alpha_s^2 M}{27\sin^22\theta_W M_Z^8r_1r_2r_3^3} 
|\tilde \Psi_{\cal V}(0)|^2 |\tilde \Psi_{\cal P}(0)|^2\sqrt{[r_3^2-(r_1-r_2)^2][r_3^2-(r_1+r_2)^2]}\times
\end{equation}
\begin{displaymath}
\Bigl[r_3^4-2r_3^2(r_1^2+r_2^2)+(r_1^2-r_2^2)^2\Bigr]F_{QG}^2.
\end{displaymath}

The numerical values of relativistic parameters $\omega_{1}$ and $\tilde{\omega}_{1}$, 
as well as relativistic wave functions at zero $\tilde{R}(0)$ and $\tilde{R}'(0)$ are given 
in Table~\ref{tb1}. We have included in this table the parameter values for some 
$S$-wave and $P$-wave states of charmonium and bottomonium, since both can be produced in pairs in decays 
of the $Z$-boson \cite{apm2025yaf}.
\begin{table}[h]
\caption{Basic parameters of some $S$- and $P$-states of charmonium and bottomonium}
\bigskip
\label{tb1}
\begin{tabular}{|c|c|c|c|c|c|c|}
\hline
Meson            & $J^{PC}$   & Mass, MeV  & $\omega_1$ & $\tilde{\omega}_1$ & 
$\tilde{R}(0) \text{ GeV}^{3/2}$ & $\tilde{R}'(0) \text{ GeV}^{5/2}$ \\ 
\hline
$J/\Psi(1S)$     & $1^{--}$   & $3096.900$ & $0.20$ & ---    & $0.81$ & ---      \\  
$\eta_c(1S)$     & $0^{-+}$   & $2984.09$   & $0.20$ & ---   & $0.92$ & ---      \\  
$\Upsilon(1S)$   & $1^{--}$   & $9460.40$  & $0.05$ & ---    & $1.88$ & ---      \\  
$\eta_b(1S)$     & $0^{-+}$   & $9398.7$   & $0.05$ & ---    & $1.95$ & ---      \\  
$\Psi(2S)$       & $1^{--}$   & $3686.097$ & $0.16$ & ---    & $0.56$ & ---      \\  
$\chi_{c0}(1P)$  & $0^{++}$   & $3415.50$  & ---    & $0.04$ & ---    & $0.33$   \\  
$\chi_{c1}(1P)$  & $1^{++}$   & $3510.67$  & ---    & $0.05$ & ---    & $0.20$   \\  
$\chi_{c2}(1P)$  & $2^{++}$   & $3556.17$  & ---    & $0.07$ & ---    & $0.13$   \\  
$h_c(1P)$        & $1^{+-}$   & $3525.37$  & ---    & $0.06$ & ---    & $0.17$   \\  
\hline
\end{tabular}
\end{table}

\begin{figure}[htbp]
\centering
\includegraphics[scale=1.]{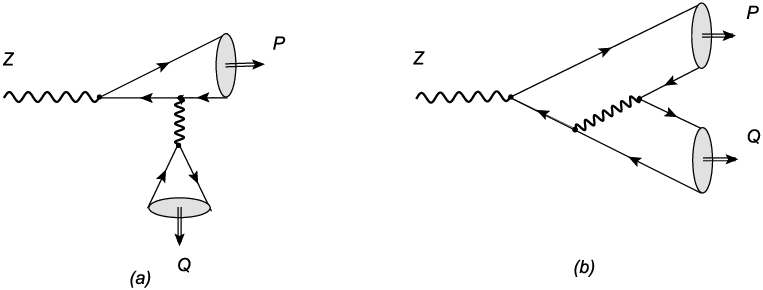}
\caption{Quark--photon mechanism of a pair quarkonium production. $P$, $Q$ are quarkonium four-momenta.}
\label{Z-Q-gamma}
\end{figure}

Let us now consider another quark-photon production mechanism, which is represented in Fig.~\ref{Z-Q-gamma} 
by two amplitudes. The second amplitude in Fig.~\ref{Z-Q-gamma}(b) is similar to the amplitudes 
in Fig.~\ref{Z-QG}, and replacing a gluon 
with a photon leads to a decrease in its contribution compared to the one considered by approximately 
10 times (replacing $\alpha_s$ to $\alpha$). The situation is different with the first amplitude in 
Fig.~\ref{Z-Q-gamma}(a), for which the same replacement of the interaction constant also takes place. 
In this amplitude, the photon emitted by a quark or antiquark is then converted into vector charmonium. Therefore, 
the square of the charmonium mass appears in the photon propagator instead of the square 
of the mass of the $Z$-boson in previous amplitudes in Fig.~\ref{Z-QG}. Thus, the contribution 
of such an amplitude increases due to the ratio $M_Z^2/M_{J/\Psi}^2$ \cite{bodwin2013,bodwin2014}.

Formally, there are four amplitudes of type in Fig.~\ref{Z-Q-gamma}(a). However, the two amplitudes 
in which the photon is converted into pseudoscalar meson vanish. The remaining two amplitudes 
have the following form:
\begin{equation}
\label{eq25}
{\cal M}^{(1)}_{Q\gamma}(P,Q)=\frac{8\pi e\alpha Q_q^2}{\sin 2\theta_W M_Z^2 M_1^2}\int\frac{d{\bf p}}{(2\pi)^3}
\int\frac{d{\bf q}}{(2\pi)^3}
Tr\left\{\Psi^{\cal P}(q,Q)\hat\Gamma(-\hat q_2-\hat P+m)\gamma^\lambda\right\}\times
\end{equation}
\begin{displaymath}
Tr\left\{\Psi^{\cal V}(p,P)\gamma^\nu\right\}D_{\nu\lambda}(P),
\end{displaymath}
\begin{equation}
\label{eq26}
{\cal M}^{(2)}_{Q\gamma}(P,Q)=\frac{8\pi e\alpha Q_q^2}{\sin 2\theta_W M_Z^2 M_1^2}\int\frac{d{\bf p}}{(2\pi)^3}
\int\frac{d{\bf q}}{(2\pi)^3}
Tr\left\{\Psi^{\cal P}(q,Q)\hat\Gamma(\hat q_1+\hat P+m)\gamma^\lambda\right\}\times
\end{equation}
\begin{displaymath}
Tr\left\{\Psi^{\cal V}(p,P)\gamma^\nu\right\}D_{\nu\lambda}(P).
\end{displaymath}

They contain the product of two traces over the Dirac factors.
After calculating them and extracting the relativistic corrections of the second order, 
the numerator of total amplitude in the case of the quark-photon mechanism can be represented as:
\begin{equation}
\label{eq27}
{\cal N}_{Q\gamma}=\varepsilon_{\alpha\beta\lambda\sigma}v_1^\alpha v_2^\beta
\varepsilon_1^\lambda\varepsilon^\sigma M F_{Q\gamma},~~~F_{Q\gamma}=(1-2a_c)\left(2r_1-2r_1\omega_{1q}+
\frac{2}{3}r_1\omega_{1p}\right).
\end{equation}

\begin{figure}[htbp]
\centering
\includegraphics[scale=1.]{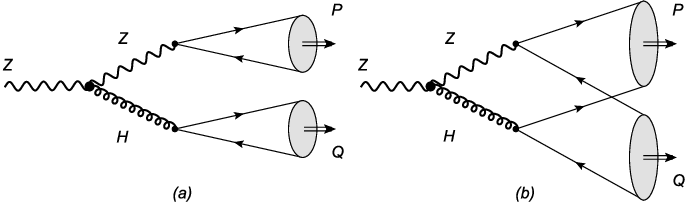}
\caption{$ZZH$ mechanism of a pair quarkonium production. $P$, $Q$ are quarkonium four-momenta.}
\label{ZZH}
\end{figure}

At tree level, there is another production mechanism, which is determined by the amplitudes 
in Fig.~\ref{ZZH} and which we call the $ZZH$ mechanism. 
A preliminary estimate of the contribution of such mechanism suggests that it can be suppressed 
by the presence of masses of the $Z$-boson and the Higgs boson in the denominator of amplitudes.
In interactions in Fig.~\ref{ZZH}, two types of new vertex factors appear:
\begin{equation}
\label{eq28}
\Gamma^{\alpha\beta}_{ZZH}=\frac{2e}{\sin 2\theta_W}M_Z g^{\alpha\beta},~~~
\Gamma_{Hq\bar q}=-\frac{e}{\sin 2\theta_W}\frac{m}{M_Z}.
\end{equation}

The direct amplitude in Fig.~\ref{ZZH}(a) will be zero regardless of which quarkonium states 
(vector or pseudoscalar) are considered.
The two cross-amplitudes are defined by following expressions:
\begin{equation}
\label{eq29}
{\cal M}^{(1)}_{ZZH}(P,Q)=-\frac{8\pi e\alpha}{\sin^3 2\theta_W}\int\frac{d{\bf p}}{(2\pi)^3}
\int\frac{d{\bf q}}{(2\pi)^3}
Tr\left\{\varepsilon^\alpha\Psi^{\cal V}(p,P)\Gamma^\beta \Psi^{\cal P}(q,Q)\right\}
D_{Z{\beta\alpha}}\left(\frac{r}{2}\right)D_H\left(\frac{r}{2}\right),
\end{equation}
\begin{equation}
\label{eq30}
{\cal M}^{(2)}_{ZZH}(P,Q)=-\frac{8\pi e\alpha}{\sin^3 2\theta_W}\int\frac{d{\bf p}}
{(2\pi)^3}\int\frac{d{\bf q}}{(2\pi)^3}
Tr\left\{\varepsilon^\alpha\Psi^{\cal P}(q,Q)\Gamma^\beta\Psi^{\cal V}(p,P)\right\}
D_{Z{\beta\alpha}}\left(\frac{r}{2}\right)D_H\left(\frac{r}{2}\right),
\end{equation}

The numerator of the sum of amplitudes \eqref{eq29}-\eqref{eq30} has the same structure 
as in previous amplitudes \eqref{eq18}, \eqref{eq27}:
\begin{equation}
\label{eq31}
{\cal N}_{ZZH}=\varepsilon_{\alpha\beta\lambda\sigma}v_1^\alpha v_2^\beta
\varepsilon_1^\lambda\varepsilon^\sigma F_{ZZH},~~~F_{ZZH}=(1-2a_c)\left(\frac{1}{2}-\frac{1}{2}\omega_{1q}-
\frac{1}{6}\omega_{1p}\right).
\end{equation}

\begin{table}[htbp]
\caption{Numerical results for relative decay widths in non-relativistic approximation 
and taking into account relativistic corrections}
\bigskip
\label{tb2}
\begin{tabular}{|c|c|c|}  \hline
Decay                         & ${\cal B}r_{nr}$    & ${\cal B}r_{rel}$  \\  \hline
$Z\to c{\bar c} g \to J/\Psi+\eta_{c}$        & $0.31\times 10^{-12}$   &  $0.11\times 10^{-12}$ \\   \hline
$Z\to b{\bar b} g \to \Upsilon+\eta_{b}$        & $5.57\times 10^{-11}$   &  $2.08\times 10^{-11}$ \\   \hline
$Z\to c{\bar c} g \to J/\Psi + J/\Psi$  & $1.01\times 10^{-12}$ &  $0.27\times 10^{-12}$ \\    \hline
$Z\to b{\bar b} g \to \Upsilon + \Upsilon$  & $1.01\times 10^{-11}$ &  $0.35\times 10^{-11}$ \\    \hline
$Z\to c{\bar c} \gamma \to J/\Psi+\eta_{c}$   & $2.90\times 10^{-11}$  &  $0.58\times 10^{-11}$  \\  \hline
$Z\to b{\bar b} \gamma \to \Upsilon+\eta_{b}$   & $5.69\times 10^{-11}$  &  $1.83\times 10^{-11}$  \\  \hline
$Z\to c{\bar c} \gamma \to J/\Psi+J/\Psi$   & $0.69\times 10^{-10}$  &  $0.16\times 10^{-10}$  \\  \hline
$Z\to b{\bar b} \gamma \to \Upsilon+\Upsilon$   & $0.82\times 10^{-12}$  &  $0.27\times 10^{-12}$  \\  \hline
$Z\to ZH\to J/\Psi+\eta_{c}$              & $5.63\times 10^{-20}$    &  $0.32\times 10^{-20}$  \\   \hline
$Z\to ZH\to \Upsilon+\eta_{b}$              & $1.04\times 10^{-17}$    &  $0.25\times 10^{-17}$  \\   \hline
$Z\to ZH\to J/\Psi+J/\Psi$                  & $0.83\times 10^{-17}$    &  $4.03\times 10^{-17}$    \\  \hline
$Z\to ZH\to \Upsilon+\Upsilon$                  & $0.64\times 10^{-15}$    &  $0.21\times 10^{-15}$    \\  \hline
\end{tabular}
\end{table}

The ratio of factors in amplitudes for different decay mechanisms is determined by the following expression:
\begin{equation}
\label{eq32}
{\cal R}_{VP}=1+\frac{3\alpha M_Z^2 Q_q^2}{16\alpha_s M^2_{1}}\frac{F_{Q\gamma}}{F_{QG}}-
\frac{\alpha M_Z^2}{8\alpha_s\sin^22\theta_W(M_H^2-\frac{M_Z^2}{4})}\frac{F_{ZZH}}{F_{QG}},
\end{equation}
when we take the factor corresponding to the quark-gluon mechanism out of the general bracket.
Table~\ref{tb2} shows the numerical values of the decay widths for individual mechanisms.
Due to the mass factor $M_Z^2/M^2$ in \eqref{eq32}, the quark-photon mechanism gives the main contribution 
to the total decay width.

So far we have considered the production of a pair of vector and pseudoscalar states of heavy quarkonium.
Proceeding in a similar manner, one can calculate the decay widths of the $Z$-boson into a pair 
of vector charmonia $J/\Psi$ or bottomonium $\Upsilon$.
The general structure of decay amplitudes has the same form as \eqref{eq4}-\eqref{eq7}, 
but a second polarization vector $\varepsilon^\sigma_2$ appears in the projection operator \eqref{eq15}, 
corresponding to the second vector quarkonium. 
By extracting the polarization vectors of three vector mesons, the decay amplitude in the case 
of quark-gluon mechanism can be represented as follows:
\begin{equation}
\label{eq31a}
{\cal M}_{QG}(Z\to V+V)=\varepsilon_{\alpha\beta\sigma\lambda}v_1^\alpha v_2^\beta \varepsilon^\lambda
[\varepsilon_1^\sigma(v_1\varepsilon_2)-\varepsilon_2^\sigma(v_2\varepsilon_1)]T^{(1)}_{QG}(r_1,r_2,r_3,\omega_1)+
\end{equation}
\begin{displaymath}
\varepsilon_{\alpha\beta\sigma\lambda}\varepsilon_1^\beta\varepsilon_2^\sigma\varepsilon^\lambda
(v_1-v_2)^\alpha T^{(2)}_{QG}(r_1,r_2,r_3,\omega_1),
\end{displaymath}
where the index $QG$ denotes, as before, the quark-gluon mechanism. The explicit form of the coefficient 
functions $T^{(i)}_{QG}(r_1,r_2,r_3,\omega_1)$ (i=1,2) is presented below.
Using \eqref{eq31}, we find the mean value of the squared amplitude modulus, which is required 
to calculate the decay width:
\begin{equation}
\label{eq31b}
{\overline{|{\cal M}_{QG}(Z\to V+V)}|^2}=\frac{1}{12}\left(r_3^2-4\right)^2
\left(r_3^2T^{(1)}_{QG}(r_1,r_2,r_3,\omega_1)-2T^{(2)}_{QG}(r_1,r_2,r_3,\omega_1)\right)^2.
\end{equation}

\begin{figure}[htbp]
\centering
\includegraphics[scale=1.]{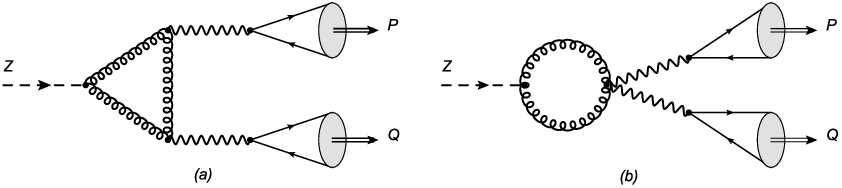}
\caption{W-loop mechanism of quarkonium pair production. $P$, $Q$ are four-momenta 
of vector quarkonium.}
\label{W-loop-Z}
\end{figure}

Since the denominators of all four amplitudes are the same (see approximation \eqref{eq16}), we also present 
here the summary expression that appears in the numerator:
\begin{equation}
\label{eq31c}
{\cal N}_{QG}(Z\to V+V)=\varepsilon_{\alpha\beta\sigma\lambda}v_1^\alpha v_2^\beta \varepsilon^\lambda
[\varepsilon_1^\sigma(v_1\varepsilon_2)-\varepsilon_2^\sigma(v_2\varepsilon_1)]
M\left(\frac{1}{2}\omega_{1q}+\frac{1}{2}\omega_{1p}\right)+
\end{equation}
\begin{displaymath}
\varepsilon_{\alpha\beta\sigma\lambda}\varepsilon_1^\beta\varepsilon_2^\sigma\varepsilon^\lambda
(v_1-v_2)^\alpha M\left(-\frac{1}{2}+\frac{1}{4}r_3^{2}\omega_{1q}+\frac{1}{4}r_3^{2}\omega_{1p}-r_q
-\frac{1}{6}\omega_{1q}-\frac{1}{6}\omega_{1p}-\frac{1}{3}r_1\omega_{1q}-\frac{1}{3}r_1\omega_{1p}\right).
\end{displaymath}

It follows from \eqref{eq31c} that in non-relativistic approximation the contribution 
to $T^{(1)}_{QG}(r_1,r_2,r_3,\omega_1)$ is equal to 0.
Let us also note that relativistic corrections to ${\cal N}_{QG}$ (second part) arise in the term 
$r_3^{2}\omega_{1q}$ that contains a large mass coefficient.

For the quark-photon mechanism we also have four production amplitudes, with the total amplitude determined 
by one term (cf. \eqref{eq31a}) in the form:
\begin{equation}
\label{eq31d}
{\cal M}_{Q\gamma}(Z\to V+V)=
\varepsilon_{\alpha\beta\sigma\lambda}\varepsilon_1^\beta\varepsilon_2^\sigma\varepsilon^\lambda
(v_1-v_2)^\alpha T^{(2)}_{Q\gamma}(r_1,r_2,r_3,\omega_1).
\end{equation}

In this case, the following numerator in the function $T^{(2)}_{Q\gamma}(r_1,r_2,r_3,\omega_1)$ is obtained:
\begin{equation}
\label{eq31e}
{\cal N}_{Q\gamma}(Z\to V+V)=
\varepsilon_{\alpha\beta\sigma\lambda}\varepsilon_1^\beta\varepsilon_2^\sigma\varepsilon^\lambda
(v_1-v_2)^\alpha M_{J/\Psi}\left(1+\frac{1}{3}\omega_{1q}+\frac{1}{3}\omega_{1p}+2r_q\right),~~~r_q=\frac{m}{M_{1}}.
\end{equation}

The average value of the square of the amplitude is determined by formula \eqref{eq31b}, in which the first 
function $T^{(1)}_{Q\gamma}(r_1,r_2,r_3,\omega_1)$ must be set equal to 0.

The ratio of factors in $T^{(2)}_{QG}(r_1,r_2,r_3,\omega_1)$ for different decay mechanisms is determined 
by the following expression:
\begin{equation}
\label{eq31f}
{\cal R}_{V V}=1+\frac{3\alpha M_Z^2 Q_q^2}{16\alpha_s M^2_{1}}
\frac{(1+\frac{2}{3}\omega_{1p}+2r_q)}{(-\frac{1}{2}+\frac{1}{2}r_3^2\omega_{1p}-r_q-\frac{1}{3}\omega_{1p}-
\frac{2}{3}r_c\omega_{1p})}-
\end{equation}
\begin{displaymath}
\frac{\alpha r_q M_Z^2}{8\alpha_s\sin^22\theta_W(M_H^2-\frac{M_Z^2}{4})}
\frac{1}{(-\frac{1}{2}+\frac{1}{2}r_3^2\omega_{1p}-r_q-\frac{1}{3}\omega_{1p}-\frac{2}{3}r_q\omega_{1p})}.
\end{displaymath}
The last term on the right-hand side of \eqref{eq31f} is determined by the $ZZH$ mechanism, the calculation 
of which is carried out using formulas \eqref{eq29}-\eqref{eq30}, taking into account the replacement 
$\gamma_5$ to $\hat\varepsilon_2$ in the projection operator \eqref{eq9}.

In our work on the decay of the Higgs boson into a quarkonium pair \cite{apm1}, 
we also considered loop decay mechanisms. 
One such mechanism, involving a loop of $W$-bosons, is shown in Fig.~\ref{W-loop-Z}.
In the case of a production of two vector quarkonium, this decay mechanism is enhanced by the appearance 
in denominators of the propagators of virtual photons the mass of produced vector meson instead 
of the mass of the $Z$-boson.
On the other side
this mechanism for producing a charmonium pair is suppressed by the power-law factor $\alpha$.

The main loop function in the amplitude shown in Fig.~\ref{W-loop-Z} is determined by two types 
of interaction vertices of the $Z$-boson with a pair of $W$-bosons and a photon with a pair of $W$-bosons 
\cite{brs,pp1}:
\begin{equation}
\label{eq33}
\Gamma_{ZWW}^{\alpha\sigma\rho}=e\cdot \cot\theta_W[g^{\sigma\rho}(2k-r)^\alpha-g^{\rho\alpha}(k+r)^\sigma+
g^{\sigma\alpha}(2r-k)^\rho],
\end{equation}
\begin{equation}
\label{eq34}
\Gamma_{AWW}^{\mu\sigma\lambda}=e [g^{\sigma\lambda}(r+Q-2k)^\mu+g^{\lambda\mu}(k+P-Q)^\sigma+
g^{\sigma\mu}(k-P-r)^\lambda],
\end{equation}
where the loop integration momentum is denoted by $k$.
When calculating the loop, different virtualities $P^2\not=Q^2$ are considered, which give different masses 
of the produced vector mesons.
The structure of the decay amplitude of the $Z$-boson into a pair of virtual photons is discussed in
\cite{matveev,pleitez}.
The loop function describing the transition of the $Z$ boson into two virtual photons can be presented 
as follows:
\begin{equation}
\label{eq35}
T^{\alpha\mu\nu}(t,P,Q)=[(P^2 r^\alpha-r^2P^\alpha)-(Q^2 r^\alpha-r^2Q^\alpha)]
[g^{\mu\nu} PQ-P^\nu Q^\mu]A(t^2,P^2,Q^2).
\end{equation}

The calculation of the function $A(t^2,P^2,Q^2)$ is carried out within the framework of the dispersion 
method \cite{t4,nishijima}.
Taking into account expression \eqref{eq35}, the total decay amplitude can be represented as:
\begin{equation}
\label{eq36}
{\cal M}(Z\to WW\to V+V)=\varepsilon_\alpha \varepsilon_{1\mu}\varepsilon_{2\nu}
\frac{(16\pi\alpha)^2}{M_1^2M_2^2}T^{\alpha\mu\nu}(t,P,Q).
\end{equation}

Calculating the imaginary parts of the amplitudes shown in Fig.~\ref{W-loop-Z} reveals that the contributions from 
the direct and crossed amplitudes (Fig.~\ref{W-loop-Z}(a)) cancel each other out exactly, resulting 
in a zero contribution 
from this mechanism to the total width. Further refinement of the decay width calculations involves 
other amplitudes, which are suppressed by powers of $\alpha$ relative to those considered here.

\section{Conclusion}

Total number of $Z$-bosons, produced (or expected to be produced) in $e^+e^-$ collisions at LEP and FCC-ee, as
well as in p-p at Tevatron, and in p-p collisions at HL-LHC, and FCC-hh 
varies in a wide range from $10^8$ to $10^{12}$ \cite{de1}.
Such a significant number of produced $Z$-bosons makes them a promising source for studying rare 
exclusive decays, the study of which is already being actively conducted at the present time 
\cite{gonsalves,CMS2023,de1,enterria}.
Among rare decays of the $Z$-boson or the Higgs boson, processes in which bound states of quarks 
or leptons are formed in the final state stand out. These processes allow us to study the dynamics 
of the formation of bound states and test the theoretical models that underlie their description.
Rare decays that produce heavy mesons and baryons offer an additional source of information about 
the interaction constants of particles in the Higgs sector. Since in such reactions the entire 
interaction process is separated into a short-range stage using the perturbative Standard Model 
and a long-range stage requiring nonperturbative QCD, obtaining new information about 
the nonperturbative interactions of many heavy quarks opens up new possibilities for studying 
tetraquarks and pentaquarks \cite{apm2,apm2025yaf}.

This work, which examines rare decays of the $Z$-boson, expands our research into the formation of bound 
states of heavy quarks, previously conducted for the Higgs boson decays. At least three important features 
can be identified for these processes. The first is the existence of different mechanisms for the production 
of mesons and baryons \cite{apm1,p2,ref5,ref6,ref7,ref8}. These mechanisms are determined by the product of the interaction constants and the mass factors of the particles participating in the reactions. It is not clear 
in advance which mechanism will be dominant, as various parameters are intertwined. The second feature 
of these processes is the significant role of relativistic effects \cite{ref1,ref2,ref3,ref4}, without which 
a reliable description is virtually impossible. Finally, the third feature is the important role of radiative 
corrections \cite{p4,p5,p6}, which contribute to the decay widths comparable to the contribution 
of relativistic corrections.

The calculation of the decay widths of the $Z$-boson with paired charmonium or bottomonium production performed 
in this paper is based on a relativistic approach using the relativistic quark model. In this approach, 
relativistic effects are determined by the momenta of the relative motion of quarks, 
are parameterized by a specific set of quantities, and can be calculated within the quark model itself. 
This requires using a Hamiltonian of the system that also takes into account the effects of the relative 
motion of heavy quarks.

The results of calculating various decay widths, presented in Table ~\ref{tb1}, demonstrate the importance 
of taking into account various decay mechanisms. Relativistic corrections, which are determined using 
the parameters $\omega_n$, $\tilde\omega_n$, and $\tilde R(0)$, significantly alter the results 
of calculations in the nonrelativistic approximation. Therefore, they must be taken into account 
in the case of the formation of bound states of heavy quarks to obtain reliable predictions 
of the decay widths. In our approach, we take into account relativistic effects connected with the law 
of transformation of meson wave functions upon transition from the rest frame to the moving reference frame, 
relativistic corrections in the interaction amplitude, and relativistic corrections when calculating 
the wave function of bound states of quarks in the rest frame using the Breit Hamiltonian.
It is useful to note that the very structure of considered amplitudes depends 
on whether relativistic corrections are taken into account. Thus, when studying the pair production 
of vector qurkonium $V$, it turns out that the decay amplitude \eqref{eq31a} contains two terms, 
with the first term vanishing in the nonrelativistic approximation. Other similar amplitudes \eqref{eq31d} 
contain only terms of one type. Another feature of relativistic description is the change in the magnitude 
of relativistic corrections in the amplitudes due to the appearance of terms of the type 
$\omega_n r_3^2$ (see, \eqref{eq31c}), in which relativistic effects are enhanced by mass factors.

In this paper, we consider four mechanisms of pair quarkonium production in $S$-states: the quark-gluon, 
quark-photon, $ZZH$ and W-loop mechanisms. 
Further study of quarkonium production in $S$-states may be connected with
the quark loop mechanism, which can be appropriately considered 
with one-loop corrections to the production amplitudes. Although this paper consider second-order corrections 
in the relative momenta of heavy quarks, the analysis can be extended to higher-order relativistic corrections. 
A distinctive feature of our work is the inclusion of relativistic corrections to nonrelativistic decay widths. 
Overall, we can say that there is order-of-magnitude agreement with previous calculations 
in the nonrelativistic approximation carried out in \cite{de1,p4}.
The analytical formulas for the widths of various decays \eqref{eq24}, \eqref{eq31b},
include many parameters, each of which contributes 
to the overall theoretical error of the calculations, since all these parameters are themselves determined 
by the calculations. Suffice it to say that even the value of strong interaction constant strongly 
depends on the choice of energy scale. But the main theoretical error in calculations is connected 
with one-loop corrections to the considered interaction amplitudes, which can be no less than 30 percent.

\begin{acknowledgments}
This work was supported by the Foundation for the Development of Theoretical Physics 
and Mathematics BASIS (grant 25-1-4-15-1) (F.A.M.).
\end{acknowledgments}

\end{document}